\documentclass[preprint,prc]{revtex4}
\usepackage{graphicx}
\usepackage[dvips,usenames]{color}
\usepackage{amsmath,amssymb}
\newcommand{\bs}{\begin{sloppypar}} \newcommand{\es}{\end{sloppypar}}
\def\beq{\begin{eqnarray}} \def\eeq{\end{eqnarray}}
\def\beqstar{\begin{eqnarray*}} \def\eeqstar{\end{eqnarray*}}
\newcommand{\bal}{\begin{align}}
\newcommand{\eal}{\end{align}}
\newcommand{\beqe}{\begin{equation}} \newcommand{\eeqe}{\end{equation}}

\begin {document}

\title{Phenomenology of Heavy Flavors in Ultrarelativistic Heavy-Ion Collisions}%
\author{ A. A. Isayev}
 \affiliation{Kharkov Institute of
Physics and Technology, Academicheskaya Street 1,
 Kharkov, 61108, Ukraine
\\
Kharkov National University, Svobody Sq., 4, Kharkov, 61077,
Ukraine
 }

\begin{abstract}
Some recent  experimental results obtained in collisions of  heavy
nuclei ($\sqrt{s}=200$ GeV) at BNL Relativistic Heavy-Ion Collider
(RHIC) are discussed.  The probes of dense matter created in
heavy-ion collision by quarkonia, $D$ and $B$ mesons containing
heavy charm and beauty quarks are considered. The centrality,
rapidity and transverse momentum dependences of the nuclear
modification factor and elliptic flow coefficient are presented
and their
possible theoretical interpretation is provided.%
\end{abstract}
 \maketitle \noindent

\section{Introduction}

 Lattice QCD (LQCD) calculations predict that at a critical temperature
  $T_c\simeq 170\,\mbox{MeV}$, corresponding to an
energy density  $\varepsilon_c \simeq 1\,\mathrm{GeV/fm^3}$,
nuclear matter undergoes a phase transition to a deconfined state
of quarks and gluons, called Quark-Gluon Plasma (QGP). At the
modern collider facilities such as CERN supersynchrotron (SPS)
(the nucleon-nucleon (NN) centre-of-mass energy for collisions of
the heaviest ions  is $\sqrt{s} =17.3\,\mbox{GeV}$), Relativistic
Heavy Ion Collider (RHIC) at Brookhaven ($\sqrt{s}
=200\,\mbox{GeV}$), and Large Hadron Collider (LHC) at CERN
($\sqrt{s} =5.5\,\mbox{TeV}$), whose heavy-ion program will start
soon, heavy-ion collisions are used to attain the energy density,
exceeding $\varepsilon_c$. This makes the QCD phase transition
potentially realizable within the reach of the laboratory
experiments. The objective is then to identify and to assess
suitable QGP signatures, allowing to study the properties of QGP.
To that end, a variety of observables (probes) can be
used~\cite{Isayev:CMS1}-\cite{Isayev:JPG2}. Further we will be
mainly interested in heavy-flavour probes of QGP, i.e., utilizing
particles having c- and b-quarks.

A special role of heavy $Q = (c, b)$ quarks as probes of the
medium created in heavy-ion collision (HIC) resides on the fact
that their masses ($m_c\approx 1.3\,\mbox{GeV},\, m_b\approx
4.2\,\mbox{GeV}$) are significantly larger than the typically
attained ambient temperatures or other nonperturbative scales,
$m_Q\gg T_c, \Lambda_{QCD}=0.2\,\mbox{GeV}$~\cite{Isayev:RH}. This
has several implications: (i) The production of heavy quarks is
essentially constrained to the early, primordial stages of  HIC.
Hence, heavy quarks can probe the properties of the dense matter
produced early in the collision. (ii)~Thermalization of heavy
quarks is "delayed"\, relative to light quarks. One could expect
that heavy quarks could "thermalize"\, to a certain extent, but
not fully on a timescale of the lifetime of the QGP. Therefore,
their spectra can be significantly modified, but still  retain
memory about their interaction history, and, hence, represent a
valuable probe. (iii) RHIC, and especially LHC experiments allow
to reach very low parton momentum fractions $x$, where gluon
saturation effects become important. Heavy quarks are useful tools
to study gluon saturation, since, due to their large masses, charm
and bottom cross sections are calculable via perturbative QCD  and
their yield is sensitive to the initial gluon density.

The heavy-flavor hadrons we will be interested in include: 1) open
charm $D=(c \bar q)$ and open beauty $B=(b \bar q)$ mesons
composed of a heavy quark $Q=(c, b)$ and a light antiquark $\bar
q=(\bar u, \bar d)$. These mesons could be sensitive to the energy
density of the medium through the mechanism of in-medium energy
loss; 2)  hidden charm [charmonia=($c \bar c$)] and hidden beauty
[bottomonia=($b \bar b$)] mesons (called collectively heavy
quarkonia) being the bound states of the charm quark-antiquark, or
bottom quark-antiquark pairs, respectively. Heavy quarkonia could
be sensitive to the initial temperature of the system through the
dissociation due to color screening of the color charge that will
be discussed later.

  For detecting heavy flavor hadrons, different decay channels
are used.  At RHIC, in PHENIX and STAR experiments  the
measurement of the spectra of open heavy flavors is based on the
measurement of the spectra of heavy flavor (HF) electrons and
positrons [($e^++e^-)/2$] from the semileptonic decays like
$D^0\rightarrow K^-e^+\nu_e$, $D^+\rightarrow \bar K^0 e^+ \nu_e$,
etc. These measurements are based on the fact that the decay
kinematics of HF electrons/positrons largely conserves the
spectral properties of the parent particles. Besides, the STAR
experiment has the capability to directly reconstruct open heavy
flavor mesons through the hadronic decay channels like
$D^0(\overline{D}^{\,0})\rightarrow K^\mp \pi^\pm$,
$D^+\rightarrow K^- \pi^+\pi^+$, etc. At low transverse momentum,
the STAR experiment also uses heavy flavor decay muons to provide
open heavy flavor measurements. Quarkonia in both STAR and PHENIX
experiments are detected through their dilepton decays $Q\bar
Q\rightarrow e^+e^-$ (midrapidity), $Q\bar Q\rightarrow
\mu^+\mu^-$ (forward rapidity).

\section{Heavy Flavor Probes of QGP: Quarkonia}

We begin the discussion of heavy flavor probes of QGP with heavy
quarkonia. The question we would like to address is: What could
happen with quarkonium yields in HIC  if QGP is really formed?

Let us note that if the deconfinement phase transition really
takes place in a dense medium created in  HIC, then a color charge
in the QGP will be screened analogously to the Debye screening of
an electric charge in the electromagnetic plasma. As a result of
color Debye screening of the heavy quark interaction in QGP, the
binding energy of a bound state decreases and one can expect that
this would lead to the suppression of quarkonium yields in HIC.
This idea was first suggested by H. Matsui and
H.~Satz~\cite{Isayev:MS} who predicted that color Debye screening
will result in the suppression of $J/\psi$ meson ($c\bar c$ in the
$^3S_1$ state, $M=3.097\, \mathrm{GeV}$) yields. After that, the
$J/\psi$ suppression  was considered as one of the key probes for
the QGP formation in heavy ion collisions. $J/\psi$ is especially
promising because of the large production cross-section and
dilepton decay channels which make it easily detectable.

However, soon it was realized that besides melting $J/\psi$ mesons
in QGP due to the screening of the color charge, there are also a
few competing mechanisms which could explain the suppression of
$J/\psi$ production in heavy-ion collisions. These mechanisms are
referred to as cold nuclear matter (CNM) effects. The first CNM
 effect is the absorption of $J/\psi$ by
nuclear fragments from colliding nuclei. Let us consider, e.g.,
the proton-nucleus collision. Once produced in the hard primary
parton processes, $J/\psi$ has to cross the length $L$ of nuclear
matter before exiting the nucleus, and, when traversing nuclear
matter, it can be absorbed by forthcoming nucleons of a nucleus.
The production cross section of $J/\psi$ in $p$-$A$ collision can
be parameterized as \begin{equation}
\sigma_{pA}^{J/\psi}=A\sigma_{pp}^{J/\psi}e^{-\sigma_{abs}^{J/\psi}\varrho
L},
\end{equation}
 where $\sigma_{pp}^{J/\psi}$ is the production cross section of
$J/\psi$ in $p$-$p$ collisions and $\sigma_{abs}^{J/\psi}$ is the
nuclear absorption cross section. From  the global fit to the data
on charmonium  production in $p$-$A$ collisions the value of
$\sigma_{abs}^{J/\psi}$ can be extracted, in particular, at SPS
(NA50 experiment) it was obtained that $\sigma_{abs}^{J/\psi}=
4.2\pm 0.5$ mb~\cite{Isayev:Al}.

The second CNM effect is related to shadowing of low momentum
partons. This means the depletion of low momentum partons in
nucleons bound in nuclei as compared to free nucleons. This effect
can be accounted for in terms of the modification of the parton
distribution functions in nucleon within the nucleus with respect
to the parton distribution functions in a free nucleon:
\begin{equation} R_i^A(x,Q^2)=\frac{f_i^A(x,Q^2)}{
f_i^N(x,Q^2)}<1,\quad i=q_v,q_{sea},g.\end{equation}  Here "i"\,
denotes valence quarks, sea quarks, and gluons, $x$ is the parton
momentum fraction, $Q^2$ is the momentum transfer squared. At high
energies, $J/\psi$s are dominantly produced through the gluon
fusion,  and the $J/\psi$ yield is therefore sensitive to gluon
shadowing. The underlying idea explaining the occurrence  of gluon
shadowing  is that the gluon density strongly rises at small $x$
to the point where gluon fusion, $gg\rightarrow g$, becomes
significant. In the case of proton-nucleus and nucleus-nucleus
collisions, where nuclei with large mass number $A$ are involved,
the nonlinear effects are enhanced by the larger density of gluons
per unit transverse area of the colliding nuclei. A direct
consequence of nuclear shadowing is the reduction of
hard-scattering cross sections in the phase-space region
characterized by small-$x$ incoming partons. For gluons, e.g.,
shadowing becomes important at $x\lesssim5\times10^{-2}$, and,
hence, is relevant for the conditions of RHIC and LHC. Note
however that the strength of the reduction is constrained by the
current experimental data only for $x\gtrsim10^{-3}$.

Let us consider the $J/\psi$ production at RHIC experiments. The
$J/\psi$ suppression can be characterized by a ratio called the
nuclear modification factor
\begin{equation}R_{AB}(p_{\,T},y)=
\frac{d^2N_{J/\psi}^{AB}/dp_{\,T}dy}{N_{coll}d^2N_{J/\psi}^{pp}/dp_{\,T}dy},\end{equation}
 obtained by normalizing the $J/\psi$ yield in $A$-$B$ nucleus-nucleus
collision  by the $J/\psi$ yield in $p$-$p$ collision at the same
energy per nucleon pair times the average number  of binary
inelastic NN collisions.  This ratio characterizes the impact of
the medium on the particle spectrum. If heavy ion collision is a
superposition of independent $N_{coll}$ inelastic NN collisions,
then $R_{AB}=1$, whereas $R_{AB}<1$ ($R_{AB}>1$) corresponds to
the case of the $J/\psi$ suppression (enhancement). As we
discussed already, at first, it is necessary to clarify the role
of  CNM effects on $J/\psi$ production. At RHIC, CNM effects are
studied in collisions of light deuteron and heavy gold nuclei,
when the energy density reached in the collision is not enough for
the formation of QGP.   At RHIC energies, shadowing of partons is
important and in the model calculations is implemented in two
shadowing schemes for the nuclear parton distribution functions,
the EKS model~\cite{Isayev:EKS} and NDSG model~\cite{Isayev:NDSG}.
The $J/\psi$ break-up cross sections obtained for two shadowing
schemes from the best fit to data are $\sigma_{\rm
breakup}=2.8^{+2.3}_{-2.1}$ mb (EKS) and $\sigma_{\rm
breakup}=2.6^{+2.2}_{-2.6}$ mb (NDSG)~\cite{Isayev:Adare} (in
Ref.~\cite{Isayev:Adare}, the term "break-up cross section"\, is
used  instead of the term "absorption cross section"). Although
these values are consistent, within large uncertainties, with the
corresponding value obtained at CERN SPS, a recent
analysis~\cite{Isayev:L}  shows that, in fact, the level of
$J/\psi$ CNM break-up significantly decreases with the collision
energy.

Let us now consider charmonium production in heavy ion collisions
at RHIC. Figs.~1,2  show the $p_T$-integrated $J/\psi$ nuclear
modification factor obtained in Au-Au collisions at RHIC/PHENIX
experiment as a function of centrality, parametrized by the number
of participating nucleons at mid- and forward rapidities,
respectively~\cite{Isayev:Adare,Isayev:A}. The PHENIX data are
shown by boxes. The $R_{AA}$ approaches unity for the peripheral
collisions (small $N_{part}$) and goes down to approximately 0.2
at most central collisions (large $N_{part}$). To see the level of
the anomalous suppression beyond the cold nuclear matter effects
it is necessary to extrapolate the CNM effects obtained in d-Au
collisions to Au-Au collisions within the given shadowing scheme
and corresponding $J/\psi$ break-up cross section. The results are
shown by black and red curves with the corresponding error bands.
It is seen that $J/\psi$ production is significantly suppressed
beyond CNM effects at forward rapidity  and suppression is less
pronounced at midrapidity in most central Au-Au collisions.

\begin{figure}[tb]
\begin{center} \includegraphics[width=8.6cm,keepaspectratio]{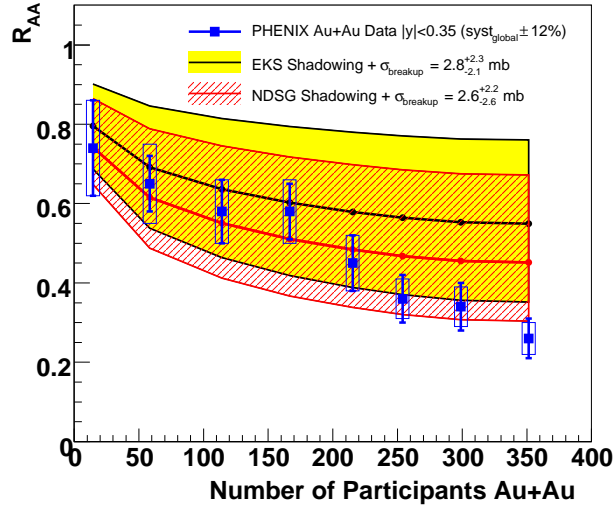}
\end{center}\vspace{-4ex}
\caption{(Color online) $J/\psi$'s $R_{AA}$ for $Au$-$Au$
collisions at midrapidity compared to a band of theoretical curves
for the breakup values found to be consistent with the $d$-$Au$
data. Both EKS and NDSG shadowing schemes are
included.}\label{Isayev:fig:0}
\end{figure} 

Thus, one can conclude that assuming the conservative cold nuclear
matter approaches some level of the anomalous suppression is
indeed observed at RHIC which could characterize the produced
medium as hot and deconfined.

\begin{figure}[tb]
\begin{center}\includegraphics[width=8.6cm,keepaspectratio]{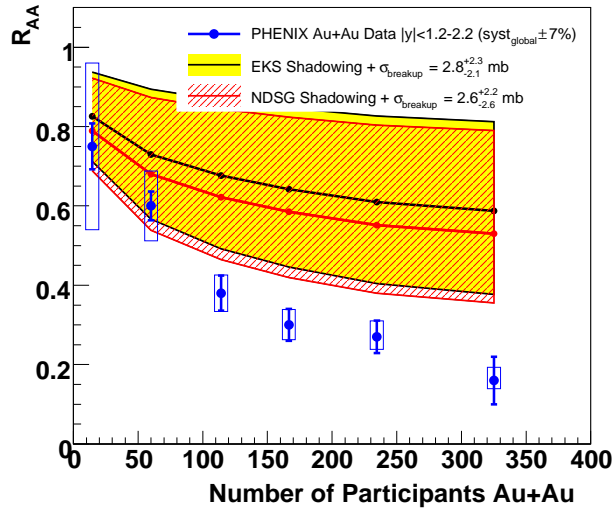}
\label{Isayev:fig:2}\end{center}\vspace{-4ex}  \caption{(Color
online)
Same as in Fig.~\ref{Isayev:fig:0}  but at forward rapidity.}%
\end{figure}

However, not all is still clear. Measurements of the $J/\psi$
suppression by PHENIX collaboration at RHIC lead to some
surprising features. Fig.~3 shows compiled data for the nuclear
modification factor obtained in CERN SPS and RHIC PHENIX
experiments. There are two surprising results in these
measurements. First, the mid rapidity suppression in PHENIX (the
red boxes) is lower than the forward rapidity suppression (blue
boxes) despite the experimental evidence that the energy density
is higher at midrapidity than at forward rapidity, and, hence, one
could expect that at midrapidity  $J/\psi$ will be more suppressed
due to higher density of color charges. Secondly, the nuclear
modification factor $R_{AA}$ at midrapidity in PHENIX (red boxes)
and SPS (black crosses) are in agreement within error bars, a
surprising result considering that the energy density reached at
RHIC is larger than the one reached at SPS. This indicates that at
RHIC energies additional mechanisms countering the suppression,
could be operative.

\begin{figure}[tb]
\begin{center} \includegraphics[width=8.6cm,keepaspectratio]{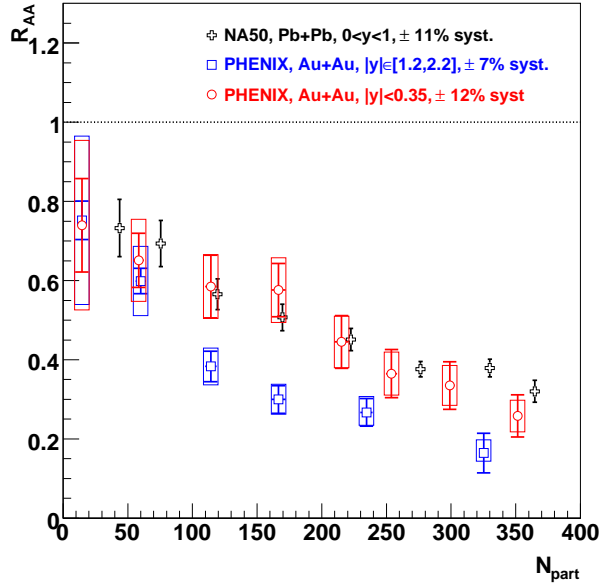}
\label{Isayev:fig:3}
\end{center}\vspace{-5ex}
\caption{(Color online) $J/\psi$ nuclear modification factor for
the most energetic SPS (Pb-Pb) and RHIC (Au-Au) collisions, as a
function of the number of participants $N_{part}$.}%
\end{figure}

Let us consider possible explanations of the above features. 1.
Regeneration of $J/\psi$ in the hot partonic phase from initially
uncorrelated $c$ and $\bar c$ quarks (quark coalescence model). If
to compare the $J/\psi$ suppression pattern at RHIC and SPS,
$J/\psi$ could be indeed more suppressed at RHIC than at SPS, but
then regenerated during (or at the boundary of) the hot partonic
phase from initially uncorrelated $c$ and $\bar c$ quarks. If to
compare the results at RHIC (midrapidity vs. forward rapidity), at
midrapidity, due to the higher energy density, there are more $c$
and $\bar c$  quarks to regenerate than at forward rapidity that
could explain the stronger suppression at forward rapidity. Note
that the total number of initial $c\bar c$ pairs is larger than 10
in the most central Au-Au collisions. Certainly, if regeneration
is important at the RHIC conditions, it will be even more
important at the LHC conditions where more than 100 $c\bar c$
pairs is expected to be produced in the central Pb-Pb collisions.
2. $J/\psi$ production could be more suppressed at forward
rapidity due to the nuclear shadowing effects, which could be more
pronounced away from midrapidity~\cite{Isayev:dC}.

One of possible experiments aimed to verify  the quark coalescence
model is the measurement of the $J/\psi$ elliptic
flow~\cite{Isayev:I}. The idea is that if charmonia were produced
by coalescence of charm quarks, they should inherit somehow  their
flow, which is known to be quite large from the open heavy flavor
measurements (see the next Section), resulting in a higher $v_2$
than in the case of the direct production of $J/\psi$ in hard
collisions. The PHENIX experiment reported a first tentative
measurement of $J/\psi$'s $v_2$ in Au-Au
collisions~\cite{Isayev:CS}. As is seen from Fig.~4, these
proof-of-principle measurements at the current level of precision
do not allow one to distinguish between models assuming various
level of regeneration (and thus elliptic flow) and much larger
statistics  is probably needed to differentiate between different
models.

\begin{figure}[tb]
\begin{center} \includegraphics[width=8.6cm]{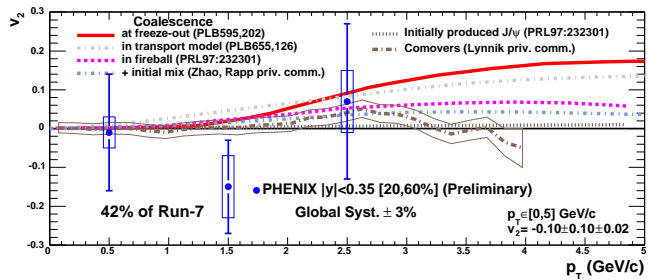}
\label{Isayev:fig:4}
\end{center}\vspace{-2ex}
\caption{(Color online) $J/\psi$'s $v_2$ at midrapidity
 for $[20, 60]$\% in centrality, as a function of $p_T$, using 42\% of 2007
Au + Au statistics, with some theoretical predictions. }%
\end{figure} 

To show the complexity of the problem, let us consider some
theoretical models for the charmonium production, which quite
satisfactory describe the RHIC data but whose predictions for LHC
are drastically different. First, in the {\it statistical
hadronization model} (SHM)~\cite{Isayev:ABRS}, it is assumed that:
1. All heavy quarks (charm and bottom) are produced in primary
hard collisions and their total number stays constant until
hadronization. 2. Heavy quarks reach thermal equilibrium in the
QGP before the chemical freeze-out (hadronization). 3.~All
quarkonia are produced (nonperturbatively) through the statistical
coalescence of heavy quarks at hadronization. Multiplicities of
various hadrons are calculated with the grand canonical ensemble.
The generation of $J/\psi$ proceeds effectively if $c, \bar c$
quarks are free to travel over large distances implying
deconfinement.

\begin{figure}
\begin{center} \includegraphics[width=8.6cm]{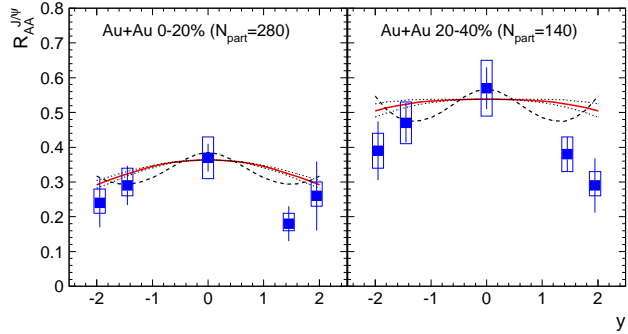}
\label{Isayev:fig:5}
\end{center}\vspace{-6ex}
\caption{(Color online) Rapidity dependence of $R_{AA}^{J/\psi}$
for two centrality classes in the statistical hadronization model.
The data from the PHENIX experiment (symbols with errors) are
compared to calculations (lines, see text).}%
\end{figure} 

Fig.~5 shows the rapidity dependence of the nuclear modification
factor, obtained in this model and the comparison with the
rapidity dependence at PHENIX for two centrality bins. Two
theoretical curves correspond to two fitting procedures, with one
and two Gaussians of the $J/\psi$ data in $pp$ collisions. In both
cases, calculations reproduce rather well (considering the
systematic errors) the $R_{AA}$ data. The model describes the
larger suppression away from midrapidity. The maximum of $R_{AA}$
at midrapidity in this model is due to the enhanced generation of
charmonium around midrapidity, determined by the rapidity
dependence of the charm production cross section. The centrality
dependence of $R_{AA}$ at $y=0$ is shown in Fig.~6. The model
reproduces quite well the decreasing trend with centrality seen in
the RHIC data. Fig.~6 also shows the prediction of the model for
the LHC. At much higher LHC energies, the charm production cross
section is expected to be larger by about an order of magnitude.
As a result, a totally opposite trend as a function of centrality
is predicted, with $R_{AA}$ exceeding unity for central
collisions.

\begin{figure}
\begin{center} \includegraphics[width=8.6cm]{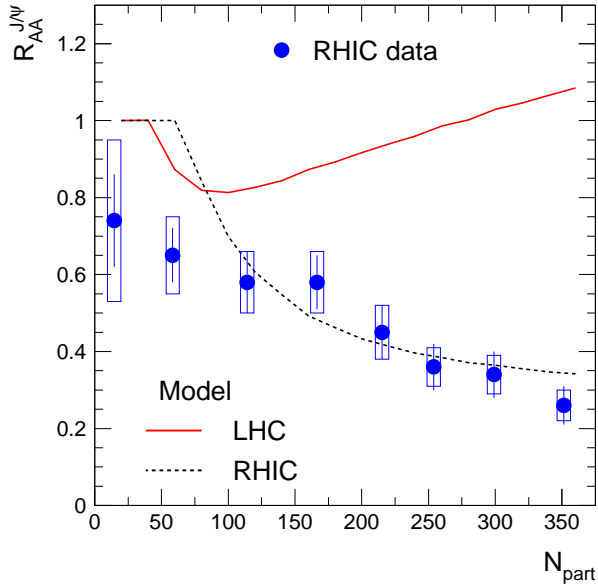}
\label{Isayev:fig:6}
\end{center}\vspace{-6ex}
\caption{(Color online) Centrality dependence of the
$R_{AA}^{J/\psi}$
 at midrapidity, according to the statistical hadronization model.}%
\end{figure}

Let us consider {\it the comovers interaction model}
(CIM)~\cite{Isayev:CBFK}. This model does not assume the
deconfinement phase transition. Anomalous suppression of $J/\psi$
(beyond CNM effects) is the result of the final state interaction
of the $c \bar c$ pair with the dense medium produced in the
collision (comovers interaction). The model consistently treats
the initial and final state effects. The initial state effects
include: 1) nuclear absorption of the pre-resonant $c \bar c$
pairs by nucleons of the colliding nuclei, 2) consistent treatment
of nuclear shadowing for hard production of charmonium. The final
state effects include absorption of the $c \bar c$ pairs by the
dense medium created in the collision (interaction with comoving
partons or hadrons produced in the collision). The model does not
assume thermodynamic equilibrium and, thus, does not use
thermodynamic concepts. The density of charmonium is governed by
the differential rate equation
\begin{equation}
\tau\frac{dn_{J/\psi}}{d\tau}=-\sigma_{co}[n_{co}(b,s,y)n_{J/\psi}(b,s,y)-\label{Isayev:2}
\end{equation}
\[\quad\quad\quad\quad\quad-n_c(b,s,y) n_{\bar c}(b,s,y)],\] supposing a pure longitudinal
expansion of the system and boost invariance. In
Eq.~(\ref{Isayev:2}), $n_{co}$ is the density of comovers, which
is found in the dual parton model~\cite{Isayev:CSTT} together with
the proper shadowing correction, $\sigma_{co}$ is the cross
section of $J/\psi$ dissociation due to interactions with
comovers, taken such as to reproduce the low energy SPS
experimental data (with $\sigma_{co} = 0.65$ mb). The first term
on the right describes dissociation of charmonium due to
interaction with comovers. The second term describes the
recombination of charmonium and is proportional to the product of
densities of charm quarks and antiquarks. The important feature of
the CIM is that recombination of $c$-$\bar c$ quarks proceeds only
locally, when the densities of quarks and antiquarks are taken at
the same transverse coordinate $s$. This is different from the
recombination in the SHM, where recombining quarks can be
separated by large distance that implies deconfinement. The
effective recombination cross section in the CIM is equal to the
dissociation cross section due to the detailed balance. The
results for the centrality dependence of the $R_{AA}$ are shown in
Fig.~7. At midrapidity, the experimental data are well reproduced
by full theoretical calculations (solid curve) taking into account
nuclear shadowing, dissociation by comovers and recombination from
charm quark and antiquark pairs. The results at forward rapidity,
presented in Fig.~\ref{Isayev:fig:8}, also well agree with the
data, in particular, the $J/\psi$ suppression at forward rapidity
is somewhat larger that the suppression at midrapidity.

\begin{figure}[tb]
\begin{center} \includegraphics[width=8.6cm,keepaspectratio]{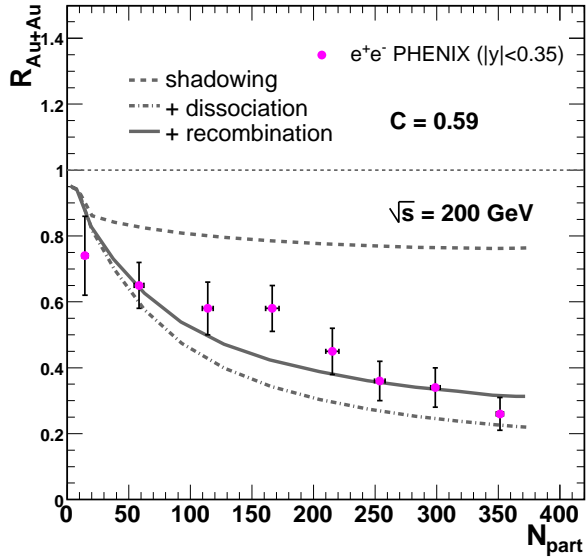}
\label{Isayev:fig:7}
\end{center}\vspace{-4ex}
\caption{Results for $J/\psi$ suppression in Au-Au collisions at
RHIC at midrapidity in the comovers interaction model. The solid
curve is the final result. The dash-dotted one is the result
without recombination ($C =
0$). }%
\end{figure} 

\begin{figure}[tb]
\begin{center} \includegraphics[width=8.6cm,keepaspectratio]{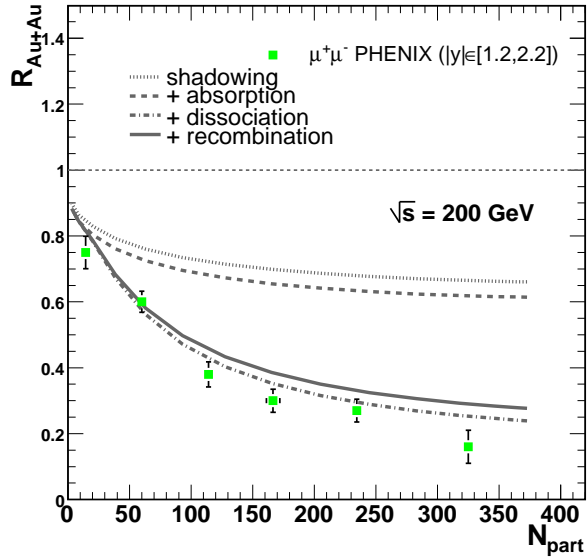}
\end{center}\vspace{-3ex}
\caption{Same as in Fig.~7, but at forward rapidity. The dashed
line is the total initial-state effect. The dotted line is the
result of shadowing. In Fig.~7,
the last two lines coincide.}\label{Isayev:fig:8}%
\end{figure} 

 Fig.~9 shows the
predictions of the model for LHC. The parameter $C$ encodes the
recombination from $c$-$\bar c$ pairs and  vanishes in the absence
of recombination. Although the density of charm grows
substantially from RHIC to LHC, the combined effect of
initial-state shadowing, absorption and comovers dissociation
overcomes the effect of parton recombination. This is in sharp
contrast with the predictions of the statistical hadronization
model where a strong enhancement of the $J/\psi$ yield with
increasing centrality was predicted.

\begin{figure}[tb]
\begin{center} \includegraphics[width=8.6cm,keepaspectratio]{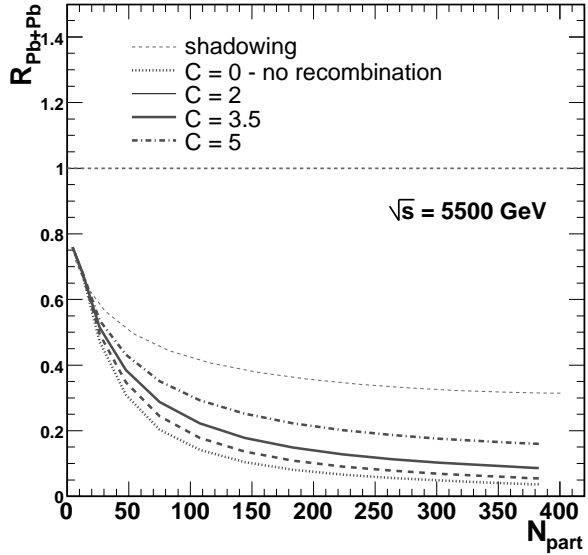}
\label{Isayev:fig:9}
\end{center}\vspace{-3ex}
\caption{Results for $J/\psi$ suppression in Pb+Pb at LHC
($\sqrt{s} = 5.5$ TeV) at
  midrapidity
  for different
values of the parameter $C$ in the comovers interaction model. The
upper line is the suppression due to initial-state effects.}%
\end{figure} 

Thus, the $J/\psi$ suppression is an important characteristic to
search for QGP. But if $J/\psi$ anomalous suppression (beyond CNM
effects) was observed in HIC, there are a few competing mechanisms
to explain that: 1. Charmonium is dissociated due to the genuine
color screening in the deconfined medium. 2. Charmonium is
dissociated through interactions with comoving partons or hadrons
in the medium formed in HIC. How to differ these mechanisms? In
comovers interaction model, the anomalous suppression sets in
smoothly from peripheral to central collisions rather than in a
sudden way when the corresponding dissociation temperature in the
deconfined medium is reached. Even at SPS, where the role of quark
recombination is of minor importance,  current experimental errors
still do not allow to disentangle these two mechanisms. However,
even if the color screening mechanism is dominating, it is unclear
what is really melted, directly produced $J/\psi$s or originating
from the feed-down of less bound charmonium states, $\chi_c$
($\rightarrow J/\psi+X$), $\psi'$ ($\rightarrow J/\psi+X$), which
have lower dissociation temperatures.  At RHIC and LHC conditions,
the feed-down from B-meson decays becomes also important.
Recombination enhances $J/\psi$ production and much complicates
the picture but its effect may be different depending on whether
the deconfinement phase transition happened or not. It is even
possible that after all $J/\psi$s were melted in QGP they can be
statistically regenerated at hadronization. True operative
mechanisms of $J/\psi$ production can be established only after
studying all important dependences (from centrality, rapidity,
collision energy $\sqrt{s}$,...) of all relevant observables with
sufficient accuracy. Further we will consider also the $J/\psi$
production in RHIC experiments as a function of the transverse
momentum, but, for the sake of comparison, this will be more
illustrative  to do after presenting the respective dependence for
open heavy flavor mesons.

So far we considered the suppression (or enhancement) patterns for
the $J/\psi$ production. Now let us briefly discuss the
perspective for bottomonia. One could expect that bottomonium
production might be easier to understand than charmonium
production due to the following reasons. 1. Since less than one
$b\bar b$ pair is produced in one central Au-Au collision, the
regeneration is negligible at the conditions of RHIC. Besides,
only about 5 $b\bar b$ pairs are expected to be produced in a
single central Pb-Pb collision at LHC. Hence, regeneration should
play much less role in the beauty sector than in the charm sector.
2. Having higher masses, bottomonia originate from higher momentum
partons and will less suffer from shadowing effects. 3. The
absorption cross section for $\Upsilon$ is by $40-50\,\%$
 smaller than the corresponding cross
section for $J/\psi$ and $\psi'$.

These features should ease the separation of the anomalous
suppression in the $\Upsilon$'s family.

\section{Heavy Flavor Probes of QGP: Open Heavy Flavor Mesons}

What qualitative effects could one expect to obtain when probing
the dense matter by heavy quarks (charm or bottom)? As  well known
from electrodynamics, the bremsstrahlung off an accelerated heavy
quark $Q$ is suppressed by the large power of its mass $\sim
(m_q/m_Q)^4$
 as compared to light quarks. Therefore,
gluon radiation off heavy quarks (i.e., radiative energy loss) is
much suppressed relative to light quarks. Consequently, one could
expect a decrease of high $p_T$ suppression and of the elliptic
flow coefficient $v_2$ from light to charm to bottom quarks. Or,
that the energy loss and coupling to matter of heavy quarks is
smaller than for light quarks as well as that the thermalization
time for heavy quarks is longer than for light quarks. Due to the
above features, one should observe a pattern of gradually
increasing $R_{AA}$ when going from the mostly gluon-originated
light-flavor hadrons ($h^\pm$ and $\pi^0$) to $D$
  to $B$ mesons: $R_{AA}^h\lesssim R_{AA}^D\lesssim
R_{AA}^B$~\cite{Isayev:Ar} (gluons lose more energy than quarks
since gluons have a higher color charge). The enhancement above
the unity of the heavy-to-light ratio
$R_{AA}^{D/h}=R_{AA}^D/R_{AA}^h$ probes the color charge
dependence of the parton energy loss while the ratio
$R_{AA}^{B/D}=R_{AA}^B/R_{AA}^D$ probes the mass dependence of the
parton energy loss.

Let us now consider what the experiment tells us about the open
heavy flavor $p_T$ suppression and elliptic flow. As mentioned
earlier, at RHIC, open heavy flavors can be  studied through the
measurements with the electrons and positrons originating from the
semileptonic decays of $D$ and $B$ mesons. Fig.~10 shows the
nuclear modification factor and elliptic flow coefficient for HF
electrons as functions of $p_T$, obtained in central collisions of
gold nuclei at RHIC (closed
circles)~\cite{Isayev:A2}-\cite{Isayev:R}. In contrast to the
above expectations, the results for $R_{AA}^{HF}$ show a strong
suppression of HF decay electrons at $p_T>2$ GeV/c, approaching at
high $p_T$ the level of suppression for $\pi^0$. This evidences
that produced medium is quite dense for heavy quarks to lose
energy as efficiently as light quarks do. The measurement of
elliptic flow gives rather large value for $v_2^{HF}$. This means
that HF electrons are involved in a collective motion being
indicative of the collective flow of their parent particles as
well.

\begin{figure}[tb]
\begin{center}\includegraphics[width=8.6cm,keepaspectratio]{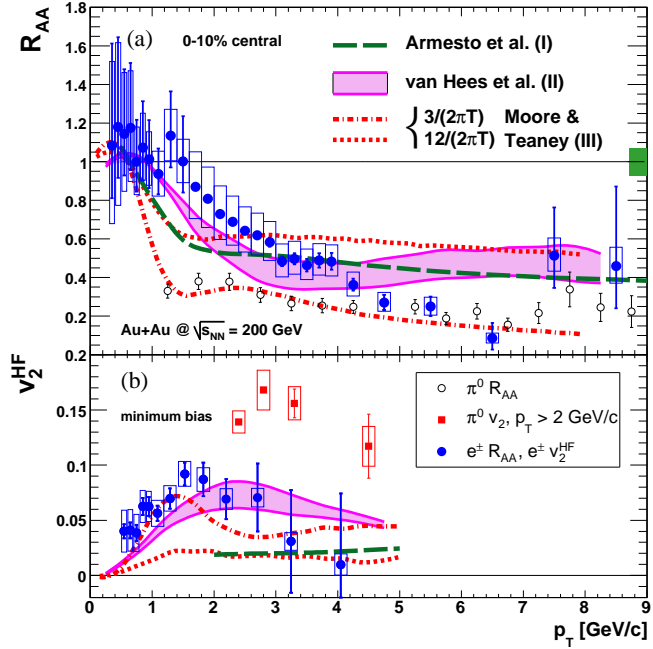}
\end{center} \vspace{-2ex} \caption{(Color
online) Nuclear modification factor (upper panel, central Au-Au)
and elliptic flow (lower panel, minimum-bias Au-Au) of
non-photonic electrons at RHIC, compared to theory. The band
corresponds to the Langevin simulations based on an expanding
fireball with effective heavy
quark resonance interactions~\cite{Isayev:HGR}.}\label{Isayev:fig:10}%
\end{figure} 

In Fig.~\ref{Isayev:fig:10}, the results of some model
considerations are also shown. The best description is provided by
the model assuming the Brownian motion of heavy quarks within the
framework of Langevin dynamics~\cite{Isayev:HGR}. Let us consider
the basic assumptions of this theory. Firstly, the thermal heavy
quark momentum $p^2\sim mT$ (in nonrelativistic approximation) is
much larger than the typical momentum transfer $Q^2\sim T^2$ from
a thermal medium to a heavy quark. Hence, the motion of a heavy
quark in the QGP can be represented as the Brownian motion, which
can be described by using the  Langevin equation. Secondly, heavy
quark loses its energy in elastic scattering processes with light
partons. Besides, as evidenced from calculations of heavy and
light meson correlators within LQCD, in QGP the $D$- and $B$-meson
like resonant states exist up to $T<2T_c$. Rescattering on these
resonant states plays an important role in thermalizing heavy
quarks. Thirdly, in order to get the spectrum of HF electrons,
$c$- and $b$-quarks are to be hadronized to $D$ and $B$ mesons via
quark coalescence (at low $p_T$) and fragmentation (at high
$p_T$).

\begin{figure}[tb]
\begin{center}  \includegraphics[width=8.6cm,keepaspectratio]{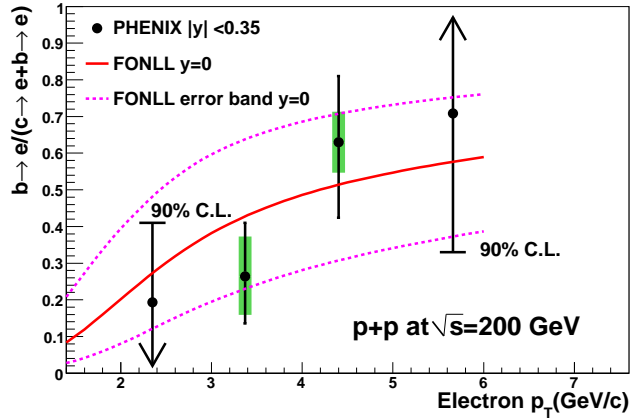}
\label{Isayev:fig:11}\end{center} \vspace{-2ex} \caption{(Color
online) Transverse momentum dependence of the relative
contribution from B mesons to the non-photonic electron yields.
The solid curve illustrates the
fixed-order-plus-next-to-leading-log (FONLL)
calculation~\cite{Isayev:FONLL}.}
\end{figure}

 The analysis shows  that resonance scattering decreases
nuclear modification factor $R_{AA}^{HF}$ and increases azimuthal
asymmetry $v^{HF}_2$. Heavy-light quark coalescence in subsequent
hadronization significantly amplifies $v^{HF}_2$ and increases
$R_{AA}^{HF}$, especially in the $p_T\simeq2$ GeV/c region. The
contribution from B mesons to $R_{AA}^{HF}$ and $v^{HF}_2$ is
estimated by providing full calculations with $c$+$b$ quarks and
with only $c$ quarks. The result is that the B-meson contribution
increases $R_{AA}^{HF}$ and decreases $v^{HF}_2$, and becomes
important above $p_T\simeq3$ GeV/c. The last point was actually
confirmed in the recent PHENIX measurements of the bottom fraction
of HF electrons, $(b\rightarrow e)/[(b\rightarrow e)+(c\rightarrow
e)]$, obtained in $p$-$p$ collisions at $\sqrt{s}
=200\,\mbox{GeV}$ and shown in Fig.~11~\cite{Isayev:bvsc}. Thus,
one can conclude, that the combined effect of coalescence of heavy
quarks $Q$ with light quarks $q$, and of the resonant heavy-quark
interaction is essential in generating strong elliptic flow
$v^{HF}_2$ of up to $10\%$, together with strong suppression of
heavy flavor electrons with $R^{HF}_{AA}$ about 0.5.

\section{Charmonium production at finite transverse momentum}

Now we can consider the production of $J/\psi$ mesons at finite
transverse momentum $p_T$ and compare it with the corresponding
dependence for open heavy flavor mesons. Fig.~12 shows the
$p_T$-dependence of $J/\psi$'s $R_{AA}$ in $0-20\%$ Cu+Cu
collisions from PHENIX~\cite{Isayev:Adare2} and
STAR~\cite{Isayev:Ab2}, and $0-60\%$ Cu+Cu collisions from
STAR~\cite{Isayev:Ab2}. The most striking peculiarity is that the
nuclear modification factor for $J/\psi$ increases with $p_T$ from
the value of about 0.5 at low $p_T$ to the value slightly
exceeding unity  at high $p_T>5\, \mathrm{GeV/c}$. These data
indicate that there is no $J/\psi$ suppression at high $p_T$ in
heavy-ion RHIC experiments~- the result which could be considered
as surprising taking into account the strong suppression of heavy
flavor electrons  considered in the previous section.

\begin{figure}[tb]
\begin{center} \includegraphics[width=8.6cm,keepaspectratio]{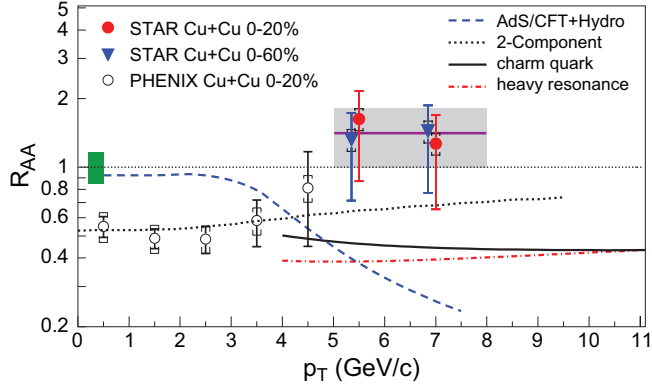}
\end{center}\vspace{-3ex} \caption{(Color
online) $J/\psi$'s $R_{AA}$ vs. $p_T$, obtained at RHIC
experiments and compared with different model calculations. The
best description of the data is provided by the two-component
model~\cite{Isayev:ZR} (dotted line).} \label{Isayev:fig:11}
\end{figure}

For comparison, Fig.~12 also shows the results of the model
calculations for the open charm $R_{AA}$. The solid line
corresponds to the charm quark energy loss because of the elastic
scattering and radiative parton processes, with assumed medium
gluon density $dN_g/dy = 254$ for $0-20\,\%$ Cu+Cu
collisions~\cite{Isayev:W}. The dash-dotted line shows the results
of the model calculations for the D-meson energy loss with
$dN_g/dy = 275$~\cite{Isayev:AV}, where the D-meson suppression is
caused by the collisional dissociation in quark-gluon plasma. Both
models, which correctly describe open heavy-flavor suppression in
Au-Au collisions, predict charm meson suppression of a factor
about 2 at $p_T > 5\,\mathrm{ GeV/c}$, in contrast to the
$J/\psi$'s $R_{AA}$. This comparison suggests that high-$p_T$
$J/\psi$ production does not dominantly proceed  via a  channel
carrying color, but rather the contribution of the color singlet
channel is prevailing.

The dotted line in Fig.~\ref{Isayev:fig:11} is the result for the
$J/\psi$'s $R_{AA}(p_T)$ in the two-component
model~\cite{Isayev:ZR}, which provides quite good qualitative
description of the data.  Let us consider the basic assumptions of
this model in more detail. First, it assumes that the $p_T$
spectra of charmonia states ($\Psi= J/\psi, \chi_c, \psi'$)
consist of two parts, direct and coalescence:

\begin{equation}
\frac{dN_\Psi}{p_Tdp_T}=\left.\frac{dN_\Psi}{p_Tdp_T}\right|_{dir}
+\left.\frac{dN_\Psi}{p_Tdp_T}\right|_{coal}.
\end{equation}

 The direct component is associated with hard
production of charmonia in primordial N-N collisions, subject to
suppression in the subsequent medium evolution due to gluon
dissociation reactions in QGP and break-up by $\pi$ and $\rho$
mesons in a hadron gas phase. The phase space distribution of
different charmonia follows the Boltzmann transport equation,
where the nuclear absorption  is included in the initial
conditions. Besides, the Cronin effect, consisting in multiple
scattering of an initial parton on elementary constituents of the
facing nucleus and leading to an increased $p_T$ in its final
state (before fusion to charmonia), is also taken into account
through initial conditions for charmonia momentum distributions.
The model consistently treats the leakage effect, i.e., charmonia
travelling outside the fireball boundary before freeze-out are not
subject to dissociation. The leakage effect reduces suppression
primarily for high-$p_T$ charmonia. The soft component is
associated with the coalescence of $c, \bar c$ quarks near the QCD
phase boundary assuming an approximate thermalization up to
$c$-quark momenta of $p_T\sim2-2.5\,\mathrm{GeV/c}$. The medium
evolution is modelled by the isentropically expanding fireball
with a cylindrical volume.

Although in such a  formulation the model is already quite
complicated, two additional effects should be taken into account
in order to reproduce an increasing trend of  $J/\psi$'s $R_{AA}$
at high $p_T$~\cite{Isayev:ZR}. These two additional effects are:
1. Finite formation time, required to build up the charmonium wave
function from a "pre-hadronic"\, $c$-$\bar c$ pair - this effect
leads to the reduction of the charmonia dissociation cross
sections. 2. Feed-down from $B$-mesons, $B \rightarrow J/\psi$.

\begin{figure}[tb]
\begin{center} \includegraphics[width=8.6cm,keepaspectratio]{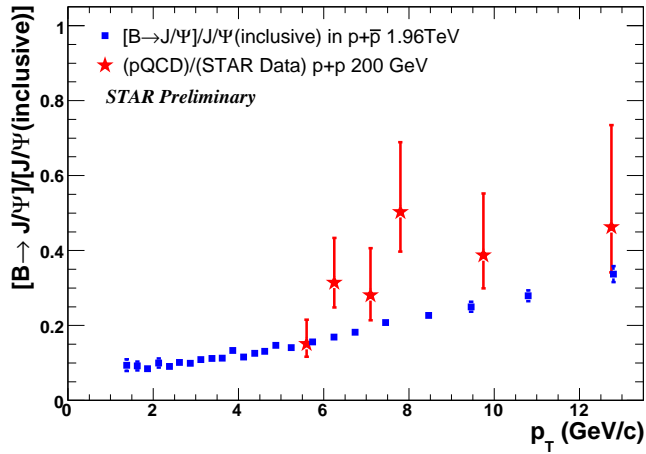}
\end{center}\vspace{-3ex} \caption{(Color
online) Tevatron and STAR data for the ratio of $J/\psi$  from
B-meson feed-down to inclusive
$J/\psi$.}\label{Isayev:fig:13}%
\end{figure} 

Fig.~\ref{Isayev:fig:13} (from Ref.~\cite{Isayev:ZR}) shows the
data from Tevatron~\cite{Isayev:Ac} and STAR on the $B\rightarrow
J/\psi$ feed-down fraction in elementary $p$-$\bar p$ ($p$-$p$)
collisions which is quite considerable at high $p_T$.  After
including the formation time effects and B-meson feed-down
(according to Tevatron data) into the two-component model, it is
able to reproduce the increasing behavior of $J/\psi$'s $R_{AA}$
at high $p_T$. Nevertheless, note that the recent STAR
measurements on $J/\psi$-hadron azimuthal correlations in $p$-$p$
collisions at $\sqrt{s}=200$~GeV show that the $J/\psi$ fraction
from B-meson feed-down at high $p_T$ is not so significant, about
$13\%$ at $p_T>5$~GeV/c. This means that further studies are
necessary, both theoretical and experimental, in order to provide
a consistent picture of $J/\psi$ production at high $p_T$.


\end{document}